\documentclass[grl]{GRLlike}
\usepackage[pdftex]{graphicx}
\usepackage{amsmath,amsfonts}

\authorrunninghead{SOLODOCH ET AL}
\titlerunninghead{Yanai-forced MJO-like signal}

\authoraddr{Aviv Solodoch, Weizmann Institute of Science, Rehovot,
  76100, Israel}

\authoraddr{
  William Boos, Zhiming Kuang, Eli Tziperman, Dept of Earth and
  Planetary Sciences and School of Engineering and Applied Sciences,
  Harvard University, 20 Oxford St, Cambridge, MA 02138-2902.
  wrboos@fas.harvard.edu, kuang@fas.harvard.edu, eli@eps.harvard.edu.}

\usepackage{lineno}

\begin{document}

\title{Excitation of slow MJO-like Kelvin waves in the equatorial
  atmosphere by Yanai wave-group via WISHE-induced convection}

\authors{Aviv Solodoch \altaffilmark{1}, William Boos
  \altaffilmark{2}, Zhiming Kuang \altaffilmark{3} and Eli Tziperman
  \altaffilmark{3}}

\altaffiltext{1}{Weizmann Institute of Science, Rehovot, 76100, Israel}

\altaffiltext{2}{Dept of Earth \& Planetary Sciences, Harvard University}

\altaffiltext{3}{Dept of Earth \& Planetary Sciences and School of
  Engineering and Applied Sciences, Harvard University}

\begin{abstract}
  The intraseasonal Madden-Julian oscillation (MJO) involves a slow
  eastward-propagating signal in the tropical atmosphere which
  significantly influences climate yet is not well understood despite
  significant theoretical and observational progress.

  We study the atmosphere's response to nonlinear ``Wind Induced
  Surface Heat Exchange'' (WISHE) forcing in the tropics using a
  simple shallow water atmospheric model.  The model produces an
  interestingly rich interannual behavior including a slow, eastward
  propagating equatorial westerly multiscale signal, not consistent
  with any free linear waves, and with MJO-like characteristics.  It
  is shown that the slow signal is due to a Kelvin wave forced by
  WISHE due to the meridional wind induced by a Yanai wave group.  The
  forced Kelvin wave has a velocity similar to the group velocity of
  the Yanai waves, allowing the two to interact nonlinearly via the
  WISHE term while slowly propagating eastward.  These results may
  have implications for observed tropical WISHE-related atmospheric
  intraseasonal phenomena.
\end{abstract}

\begin{article}

\section{Introduction}
\label{sec:Intro}

The ``Madden Julian Oscillation'' (MJO)
\citep{Madden-Julian-1971:detection, Madden-Julian-1994:observations,
  Zhang-2005:MJOrev} is the most significant intraseasonal phenomena
in the tropical atmosphere.  It couples convection and large-scale
circulation, and propagates eastward at about 5 ms$^{-1}$, through the
Indian and East to Central Pacific oceans.  This is a surprisingly
slow propagation when compared with a typical moist Kelvin wave speed
(15-20 ms$^{-1}$).

One of the mechanisms suggested to explain the MJO was wind induced
surface heat exchange (WISHE) \citep{Emanuel-1987:air,
  Neelin-Held-Cook-1987:evaporation}, according to which air-sea
latent heat fluxes enhanced by surface winds are of paramount
importance for forcing and organizing tropical convection.
\citet{Emanuel-1987:air} and \citet{Neelin-Held-Cook-1987:evaporation}
found Kelvin-like solutions due to the convection-wind coupling in
linear WISHE models.  While these early solutions seem to differ
significantly from observed MJO characteristics
\citep[e.g.][]{Zhang-2005:MJOrev}, recent observational and model
evidence suggests an important role for WISHE in MJO dynamics
\citep{Maloney-Sobel-2004:surface,
  Sobel-Maloney-Bellon-Frierson-2008:role}.  In addition, nonlinear
WISHE effects seem to be important in simulated MJO-like disturbances
\citep{Raymond-2001:new, Maloney-Sobel-2004:surface}.

In light of these findings we investigate an equivalent-barotropic
shallow water atmospheric model with a simple nonlinear WISHE
formulation, on a beta plane.  The numerical experiments yield
intriguing dynamics, including a slow eastward propagating signal that
shares some of the observed MJO characteristics, and its mechanism is
analyzed below and also presented in the thesis of
\citet{Solodoch-2010:excitation} (hereafter S2010).  Given the
simplicity of the model, it is not possible to conclusively
demonstrate that this mechanism is applicable to the observed MJO, but
we feel it is nevertheless a most interesting and nontrivial mechanism
which can enrich our understanding of tropical dynamics.

\section{The model}
\label{sec:model}

The model equations are the shallow water equations with a WISHE
heating forcing term $(Q)$ \citep{Gill-1980:some},
\begin{linenomath*}
\begin{eqnarray}
  u_t-\beta yv &=&-\eta_x + \nu\nabla^2 u,
  \label{eq:SW-u-momentum} \\
  v_t+\beta yu &=&-\eta_y + \nu\nabla^2 v,
  \label{eq:SW-v-momentum} \\
  \eta_t+H(u_x+v_y) &=&-Q(u,v,\eta) -\frac{1}{\tau}\eta.
  \label{eq:SW-continuity} \\
  Q&=&\max\left(0
    ,A\left|\mathbf{u}+\mathbf{U}\right|(\eta-\eta_{sat})\right)
  \label{eq:Q-heating}
\end{eqnarray}
\end{linenomath*}
where $u$ and $v$ are the zonal and meridional perturbation
velocities, $\eta$ is the perturbation (hydrostatic) pressure, $\nu$
the eddy diffusivity coefficient, and $\tau$ a Newtonian cooling time
scale.  In the standard experiment, mean surface easterlies,
$\mathbf{U}=U\hat{x}$, which tend to prevail in much of the equatorial
Pacific are added to the model velocity in the WISHE term only, with
$U=-2$ m/s.  We also set $\tau=5$ days, $\eta_{sat}=-5$ mbar and
$A=2\cdot10^{-6}$ m$^{-1}$.  The equivalent depth is set to $H=50$ m,
consistent with observed convectively coupled waves in the tropics
having $H=10-60$ m \citep{Wheeler-Kiladis-1999:convectively}.  The
typical tropical balance between convective heating and adiabatic
cooling is thus implicit in
(\ref{eq:SW-u-momentum})-(\ref{eq:SW-continuity}) through this choice
of $H$, and $Q$ represents only the positive-definite convective
heating caused by surface heat flux variations. The model is periodic
in the east-west coordinate ($x$), extends from 60S to 60N and 25,600
km in the east-west direction, and is based on the MIT GCM
\citep{Marshall-et-al-1997:finite, Marshall-et-al:1997:hydrostatic}.

\section{Results: slow eastward propagation of equatorial westerlies}
\label{sec:results}

\begin{figure}[tbhp]
  \begin{center}
    \includegraphics[width=7cm]{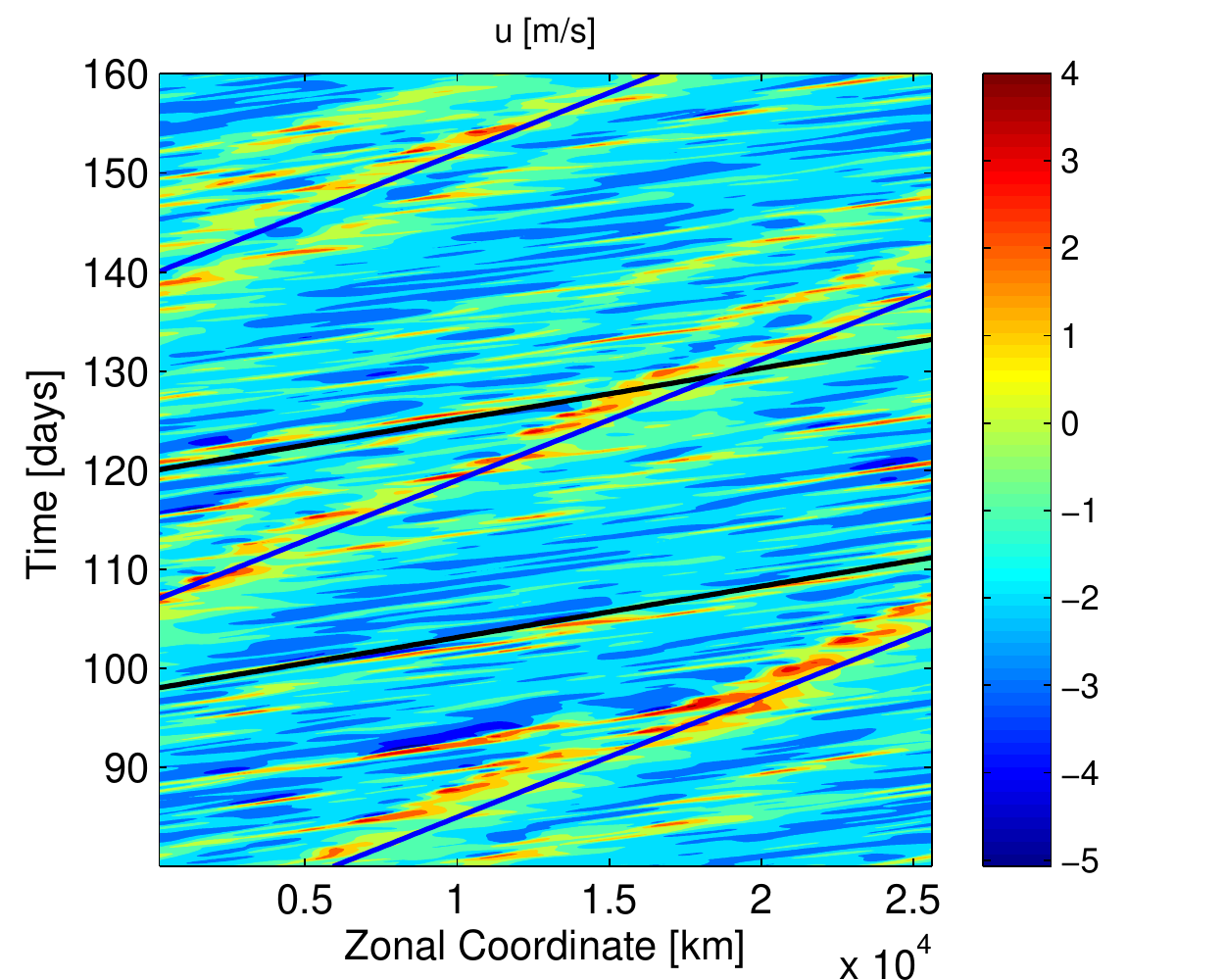}\\
    \includegraphics[width=7cm]{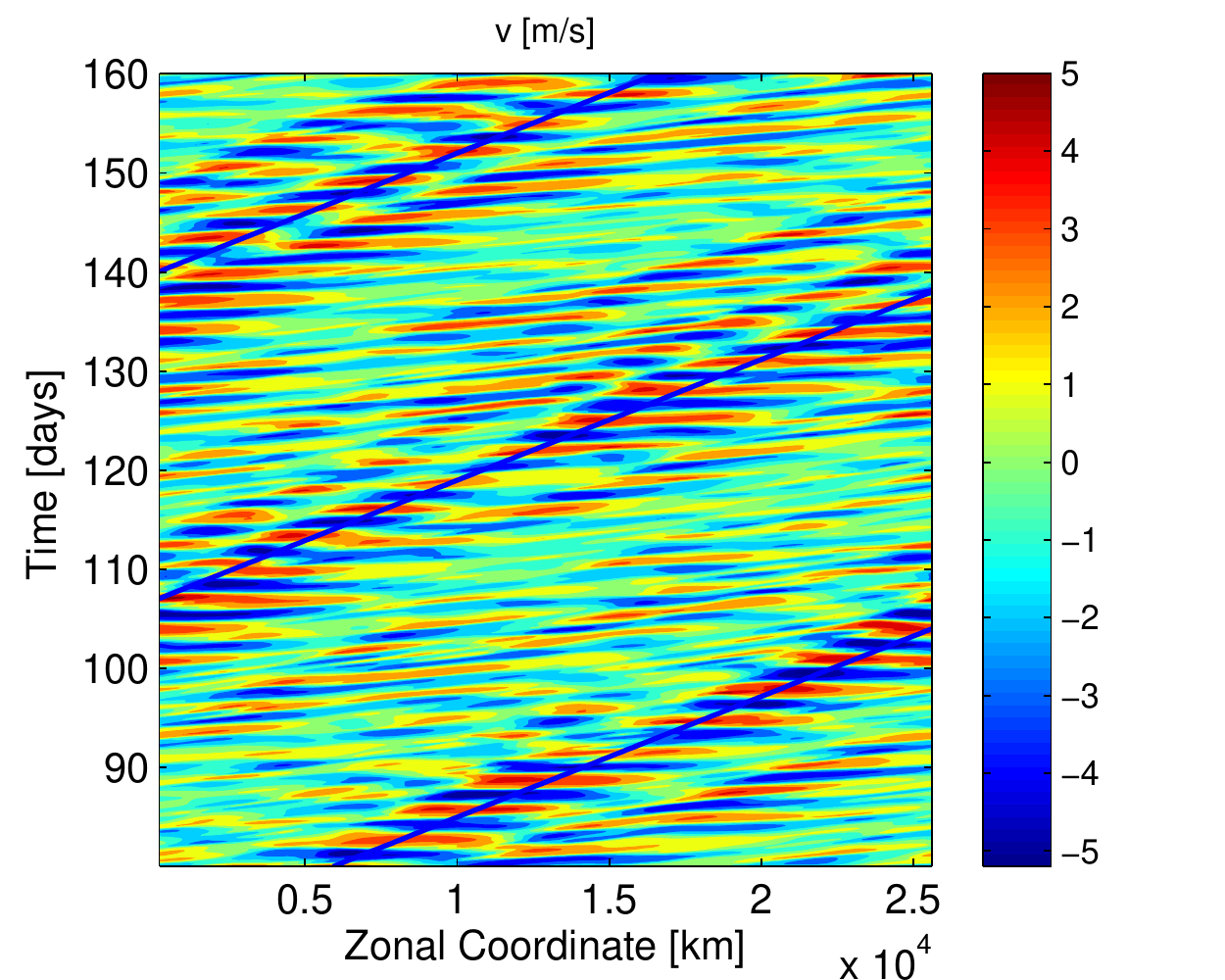}\\
    \includegraphics[width=7cm]{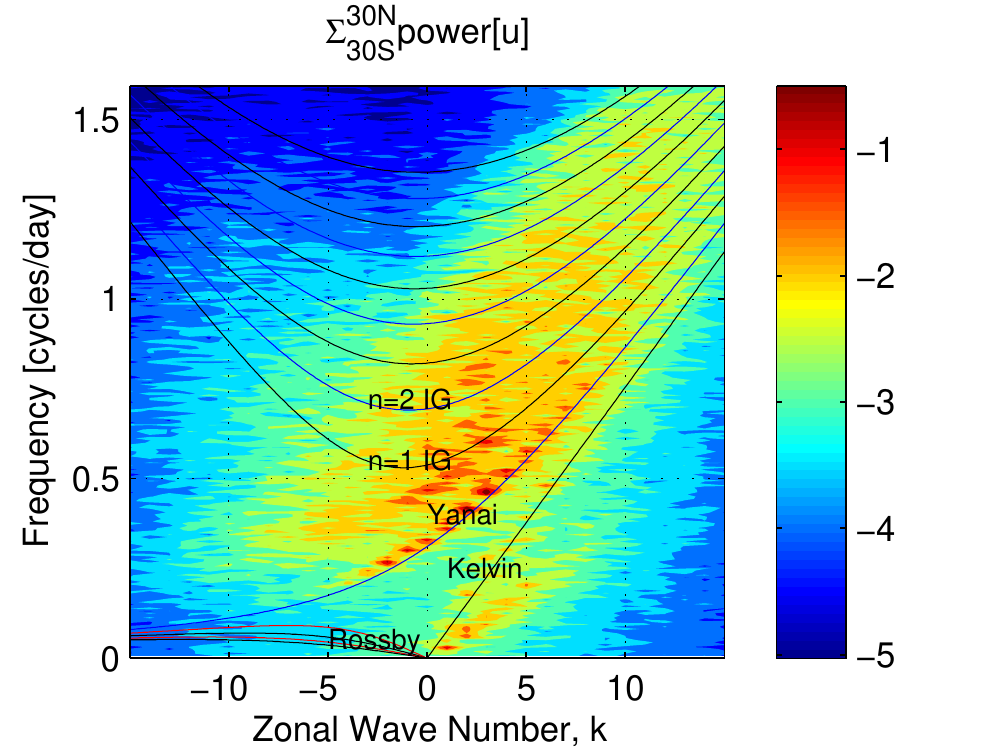}\\
  \end{center}
  \caption{Hovmoller diagrams at the equator from exp 1.  Top: zonal
    velocity, $u$ [m/s], with lines showing propagation speeds of 22
    m/s and 9.5 m/s.  Middle: meridional velocity, $v$ [m/s].  Bottom:
    Spectral power of $u(x,y,t)$, in log scale, calculated by Fourier
    transforming in $(x,t)$ at each latitude, and then summing the
    power between $\pm30$ degrees latitude.  Zonal wave number is for
    an equatorial circumference of 25,600 km.}
  \label{fig:HM-and-spectrum}
\end{figure}

A Hovmoller diagram of the zonal velocity $u$ at the equator, for days
80-160 after the initialization, is shown in
Fig.~\ref{fig:HM-and-spectrum}a.  The most prominent feature is an
eastward propagating disturbance, significantly slower than the
equatorial Kelvin wave (22 m/s in this model).  The slow signal is an
irregular, a wide band, multiscale phenomenon in space and time, and
the propagation speed varies between 8-12 m/s.  One wonders, of
course, if this slow eastward propagating equatorial disturbance is of
any relevance to the MJO, but understanding its dynamics seems a
worthwhile goal in any case.

Interestingly, the anomalous zonal wind is not symmetric, and a
histogram analysis (not shown) shows that easterly anomalies are
usually weaker than the westerly anomalies.  This is reminiscent of
``Westerly wind bursts'' observations over the tropical oceans
\citep{Harrison-Vecchi-1997:westerly} associated in some cases with
MJO events.

A Hovmoller diagram of the equatorial meridional velocity, $v$,
(Fig.~\ref{fig:HM-and-spectrum}b) shows an eastward propagating wave
group, propagating with the same speed as the slow signal seen in the
$u$ Hovmoller diagram.  The $v$ wave group envelope is highly coherent
with the slowly eastward propagating $u$ signal, while showing complex
patterns of phase propagation within the eastward propagating
envelope.  The $v$ Hovmoller and the spectrum shown in
Fig.~\ref{fig:HM-and-spectrum}c show that the $v$ signal corresponds
to a group of Yanai waves propagating with the slow signal.  It will
be shown below that the speed of propagation of the slow signal is
indeed compatible with the group velocity of long Yanai waves, and the
interaction between the $u$ and $v$ fields will be explained.

Dispersion curves of equatorial waves are superimposed over the
contours of the power spectrum in Fig.~\ref{fig:HM-and-spectrum}c, and
most of the spectral power is seen to be concentrated near these
dispersion curves.  The linear wave modes with the most power are the
Yanai wave and the Inertia-Gravity $n=1$ wave, and the Kelvin wave is
also excited.

\begin{figure}[tbhp]
  \begin{center}
    \includegraphics[width=7cm]{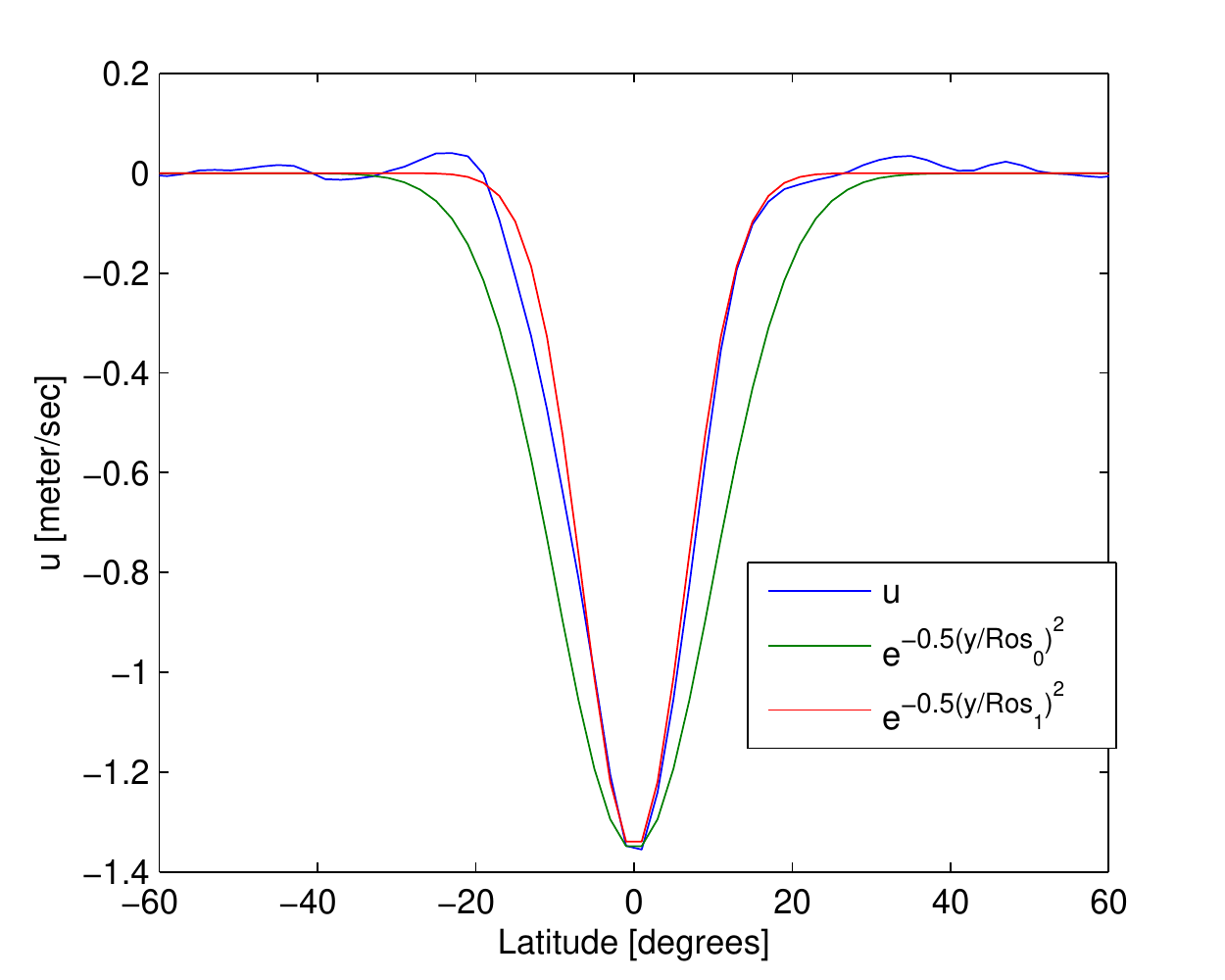}
  \end{center}
  \caption{The meridional structure of the slowly propagating signal.
    The blue line shows a meridional cross section showing the root
    time-mean square of zonal velocity anomalies at the middle
    meridian of the domain.  Green: a Gaussian meridional structure of
    the Kelvin or $v$ velocity of the Yanai wave.  Red: a Gaussian fit
    based on a Rossby radius of deformation smaller by a factor of
    $\sqrt{2}$ than the model value.}
  \label{fig:filtered-Hovmollder-and-section}
\end{figure}

However, significant power is concentrated in a feature not associated
with any linear wave mode, which is located in $k-\omega$ space
beneath the dispersion curve of the Kelvin wave.  This feature
corresponds to the slowly propagating signal seen in the Hovmoller
diagrams.  A cross section along the slowly propagating signal
(Fig.~\ref{fig:filtered-Hovmollder-and-section}, blue line) shows its
variance to be symmetric in latitude with a Gaussian structure
centered around the equator, similarly to a Kelvin wave.  It also lies
close to a linear line in $k-\omega$ space, and appears in the $\eta$
field, yet is virtually non-existent in the $v$ power spectrum
(S2010).  These are all characteristics of a Kelvin wave, except that
this feature has the ``wrong'' meridional scale when compared to the
Gaussian Kelvin structure given the model's equivalent depth
(Fig.~\ref{fig:filtered-Hovmollder-and-section}) and this is further
explained below.  We also note that most of the energy of the
``sub-Kelvin'' signal is in the $k=1$ planetary scale, and the energy
decreases with $k$.  The sub-Kelvin variability explains 50\% of the
zonal wind variability at the equator.  The above description makes it
clear that the slow eastward propagating signal clearly combines both
low frequency Kelvin-like waves, and higher frequency Yanai waves.

\section{The mechanism}
\label{sec:mech}

We now show that the slow eastward propagating signal is a forced
Kelvin wave, excited by the envelope of a Yanai wave-group.  The
efficient excitation is possible due to two factors.  First, the
forced Kelvin velocity is very similar to the Yanai group velocity.
Second, the meridional structure of the slow Kelvin waves signal is
identical to that of the square of the Yanai wave, which allows the
Kelvin wave to be excited through the nonlinear WISHE heating term.

Starting from the heating forcing term, note that the power series
expansion of the WISHE term $Q$ is,
\begin{linenomath*}
\begin{eqnarray}
  &&(\eta+|\eta^*|)\sqrt{(u+U)^2+v^2}=|\eta^*||U|+\eta|U| -|\eta^*|u 
  \nonumber \\
  &-& \eta u + \frac{|\eta^*|}{2|U|}v^2 
  +\frac{1}{2|U|}{\eta}v^2+\frac{|\eta^*|}{2|U|^2}uv^2+...
  \nonumber
\end{eqnarray}
\end{linenomath*}
Although the formal convergence criterion for this power series is not
always satisfied in the model, a correlation analysis of the different
terms in this expansion with the full heating $Q$ shows that the $v^2$
term accounts for much of the heating variability.  $v$ itself is, in
turn, dominated by the contribution due to the Yanai waves.  Now, we
are interested in the slow forcing of a Kelvin wave by the meridional
velocity due to the Yanai waves, $v_{Yanai}$, leading us to examine
the time average of $(v_{Yanai})^2$.  This represents the energy of
the Yanai waves which therefore moves at the Yanai group velocity
$c_g(k)$ and may be written as
${\langle}v_{Yanai}^2\rangle=\sum_k\frac{1}{2}\left((\widehat{v_Y^2})_ke^{ik(x-c_gt)}+c.c.\right)$,
where c.c.~is complex conjugate, and $(\widehat{v_Y^2})_k$ is the
amplitude of the envelope of $v^2$ due to the Yanai waves at
wavenumber $k$.

An important insight here is that the Yanai $v$ field has a Gaussian
structure, $\Psi_0=\exp({-{y^2}/({2R_0^2})})$, so that its square
results in WISHE forcing of the form
$Q{\sim}v^2\sim\exp({-2{y^2}/({2R_0^2})})=\exp({-{y^2}/({2R_1^2})})$,
where $R_1=R_0/{\sqrt{2}}$.  This also implies an equivalent depth of
the forced signal of $H/4$ where $H$ is the model's specified
equivalent depth.  In other words, the square of the Yanai wave $v$
velocity has the Gaussian structure corresponding to a smaller
equivalent depth, forcing a Kelvin wave like response with a similarly
smaller effective equivalent depth. This implies that this forced
Kelvin wave propagates slower than the free Kelvin waves in the
system, leading to the observed slowly propagating signal.

We found in the numerical simulation above that $Q$ excites a
Kelvin-like signal with a very small $v$ velocity.  Consider therefore
the following set of equations for a forced Kelvin wave by setting
$v=0$ and ignoring dissipation,
\begin{linenomath*}
\begin{eqnarray}
  u_t &=&-\eta_x,
  \label{eq:subKelvX} \\
  \beta yu &=&-\eta_y,
  \label{eq:subKelvY} \\
  \eta_t+Hu_x &=&
  -G(\widehat{v_Y^2})_ke^{-2\frac{y^2}{2R^2}}e^{ik(x-c_gt)}.
  \label{eq:subKelvCont} 
\end{eqnarray}
\end{linenomath*}
Assuming $G(\widehat{v_Y^2})_k$ to be a constant amplitude of the
forcing at wavenumber $k$ (see discussion below), the forced Kelvin
solution would be a superposition of the solutions with the same
functional form as the forcing,
\begin{linenomath*}
\begin{equation}
  (u,\eta)=e^{-\frac{y^2}{2R_1^2}}e^{ik(x-c_gt)}(u_0,\eta_0).
  \label{eq:forcedResponse}
\end{equation}
\end{linenomath*}
Inserting \eqref{eq:forcedResponse} into
(\ref{eq:subKelvX}-\ref{eq:subKelvY}) leads to the condition
$c_g={\beta}R_1^2={c_0}/{2}$ where $c_0$ is the free Kelvin wave speed
in the model.  Interestingly, this condition is satisfied exactly for
a Yanai wave of $k=0$ as can be verified from the Yanai dispersion
relation, and more generally is approximately satisfied for long Yanai
waves of $-5{\le}k\le10$.  This interesting coincidence allows the
slowly propagating solution \eqref{eq:forcedResponse} to be excited by
the Yanai waves.  Inserting this relation into \eqref{eq:subKelvCont}
yields the forced wave solution,
\begin{linenomath*}
\begin{equation}
  u_0 =\frac{iG(\hat{v_Y^2})_k}{k(c_0^2-c_g^2)},
  \quad
  \eta_0 = c_gu_0,
  \quad
  v = 0.
  \label{eq:subKelv_u_eta_v}
\end{equation}
\end{linenomath*}
These expressions are found to be consistent to a very good
approximation with the numerical simulation.  The disturbance clearly
has the strongest response at the longest wavenumbers ($k=1$), as is
actually found for the sub-Kelvin spectrum in our simulation (and,
interestingly, also for the observed MJO).
Fig.~\ref{fig:filtered-Hovmollder-and-section} shows a meridional
cross section through the slowly propagating signal (blue),
demonstrating that its structure is indeed very nearly a Gaussian
(red) corresponding to the square of the Yanai wave structure (green).
This provides an explicit and strong evidence of the nonlinear forcing
mechanism described above.

\begin{figure*}[tbhp]
  \begin{center}
    \includegraphics[width=18cm]{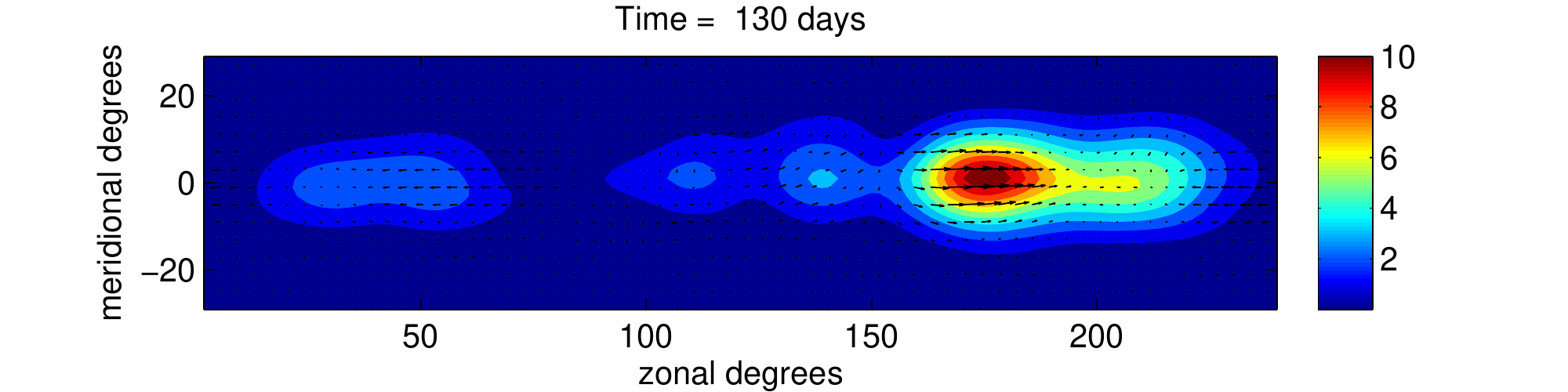}
  \end{center}
  \caption{Demonstrating the longitudinal phase between the heating
    and the slow signal wind response.  Arrows: winds corresponding to
    the slowly propagating signal (obtained by filtering the full
    solution in $k,\omega$ space, keeping only the area under the
    kelvin wave dispersion curve corresponding to the show signal
    alone.  Contours: the 2-day running time-averaged Yanai wave
    contribution to the heating ${\langle}v_{Yanai}^2{\rangle}$.}
  \label{fig:horizontal-structure}
\end{figure*}

Observations show that the MJO convection is mostly co-located with
its surface westerlies in the west-Pacific and with surface
convergence in the Indian ocean \citep[][]{Zhang-2005:MJOrev}.  The
Hovmoller plot of $u$ and $v$ (Fig.~\ref{fig:HM-and-spectrum}a,b) show
that the maximum of the Yanai envelope (and thereby its contribution
to WISHE) is nearly co-located with the maximum of the sub-Kelvin
signal at all times.  Similarly, the snapshot of
Fig.~\ref{fig:horizontal-structure} shows the maximum Yanai induced
$v^2$ WISHE envelope to be co-located with, or slightly lead, the
maximum sub-Kelvin westerlies, similar to the corresponding
relationship between MJO convection and circulation, as observed.

The Yanai waves which force the slow signal may themselves be
destabilized by the WISHE term in the presence of easterly winds
\citep{Emanuel-1993:effect}.  However, the Yanai fields on the left
hand side of the continuity equation \eqref{eq:SW-continuity} ($\eta$,
$u_x$, $v_y$) are antisymmetric in $y$
($e^{-\frac{1}{2}({y}/{R})^2}2{y}/{R}$) and therefore cannot be forced
by the symmetric, Gaussian, Kelvin wave fields ($u$ and $\eta$) even
when raised to some power, justifying the above approximation of
$G(\widehat{v_Y^2})_k$ as a constant amplitude of the forcing not
affected by the Kelvin waves.

\section{Sensitivity study}
\label{sec:sens}

The main features of the slowly propagating signal are found to be
robust with respect to different meridional boundary condition,
initial conditions, and the value of the mean parametrized zonal
velocity, $U$ (as long as it is easterly).  The signal also appears if
the WISHE forcing term is limited to the equatorial domain only, as
long as this forcing domain is not smaller than about 6 degrees
latitude.
The slow signal is also quite robust to the values of model
parameters, yet does not appear, of course, for large values of the
dissipation parameters or too small values of the WISHE coefficient.
Given that the WISHE coefficient, $A$, idealizes the product of a
transfer coefficient and several other poorly constrained
thermodynamic factors in a bulk transfer formula
\citep[e.g.,]{Emanuel-1987:air, Neelin-Held-Cook-1987:evaporation}, it
is not obvious whether the value used here is realistic.  Additional
sensitivity experiments and analysis is given in S2010.

\section{Conclusions}
\label{sec:conc}

An equivalent-barotropic shallow water model driven by nonlinear WISHE
convective heating was used to study tropical atmosphere
intraseasonal phenomena.  A slowly eastward propagating signal
appears robustly in the experiments, with about half the speed of a
linear Kelvin wave.  The slowly eastward propagating equatorial signal
is interesting because it shares some major characteristics with the
large scale fields of the MJO, some of which are not explained by current
theories.

The slow eastward propagating signal was shown to be a forced Kelvin
wave with a meridional spatial structure consistent with an equivalent
depth smaller than that specified in the model.  The signal is also
coherent with a group of long Yanai waves which has a group velocity
similar to that of the slow forced Kelvin wave.  We showed that the
Yanai group nonlinearly forces the slow Kelvin-like signal through the
WISHE term, and explained how the nonlinear WISHE term leads to the
smaller effective equivalent depth and slow propagation of the signal.
The suggested mechanism predicts the slower Kelvin wave meridional
structure and speed, as well as the right order of magnitude of the
amplitude.  The excited signal in the numerical simulations is
strongest at $k=1$ and 2, again consistent with the suggested
mechanism.  The phenomenon has a multiscale character in space and
time, and an interesting and novel aspect of the mechanism proposed
here is the critical role of higher frequency motions involving the
Yanai waves.  Concurrently with Solodoch [2010] and the present study,
a somewhat related idea has recently been presented in a conference
abstract \citep{Yang-Ingersoll-2009:testing}.

The slow signal is reminiscent of some of the main features of the
MJO, in particular its wide-band nature, slow eastward propagation,
its Kelvin-like large scale structure at and east of its convection
center, and its spectral power peaks at the $k=1$ and $2$ wavenumbers.
Additionally, there is asymmetry in zonal wind toward higher
westerlies than easterlies, particularly close to the convection
center.

Preliminary results to be reported elsewhere
\citep{Solodoch-Boos-Kuang-et-al-2010:mjo} show that the mechanism
found here is also relevant to more detailed models, with
significantly more realistic convective parameterizations.  Although
it is clearly premature to propose that the mechanism found here is
applicable to the MJO, exploring the suggested mechanism in more
realistic models and in observations may prove fruitful.

\section*{Acknowledgments}

This paper is based on the MA thesis of AS submitted to the Weizmann
Institute of Science.  ET thanks the Weizmann Institute for its
hospitality during parts of this work.  ET and ZK are funded by the
NSF climate dynamics program, grant ATM-0754332, WB acknowledges the
support of the Harvard EPS Daly postdoctoral fellowship and the
Harvard Center for the Environment fellowship.

\end{article}

\begin{thebibliography}{17}
\providecommand{\natexlab}[1]{#1}
\expandafter\ifx\csname urlstyle\endcsname\relax
  \providecommand{\doi}[1]{doi:\discretionary{}{}{}#1}\else
  \providecommand{\doi}{doi:\discretionary{}{}{}\begingroup
  \urlstyle{rm}\Url}\fi

\bibitem[{\textit{Emanuel}(1987)}]{Emanuel-1987:air}
Emanuel, K.~A. (1987), {An air-sea interaction model of intraseasonal
  oscillations in the tropics}, \textit{J. Atmos. Sci.}, \textit{44},
  2324--2340.

\bibitem[{\textit{Emanuel}(1993)}]{Emanuel-1993:effect}
Emanuel, K.~A. (1993), {The effect of convective response time on WISHE modes},
  \textit{J. Atmos. Sci.}, \textit{50}, 1763--1775.

\bibitem[{\textit{Gill}(1980)}]{Gill-1980:some}
Gill, A.~E. (1980), Some simple solutions for heat-induced tropical
  circulation, \textit{Q. J. R. Meteorol. Soc.}, \textit{106}, 447--462.

\bibitem[{\textit{Harrison and Vecchi}(1997)}]{Harrison-Vecchi-1997:westerly}
Harrison, D.~E., and G.~A. Vecchi (1997), Westerly wind events in the tropical
  {Pacific}, 1986-95, \textit{J. Climate}, \textit{10}(12), 3131--3156.

\bibitem[{\textit{Madden and Julian}(1971)}]{Madden-Julian-1971:detection}
Madden, R.~A., and P.~R. Julian (1971), Detection of a 40-50 day oscillation in
  zonal wind in tropical {Pacific}, \textit{J. Atmos. Sci.}, \textit{28}(5),
  702--\&.

\bibitem[{\textit{Madden and Julian}(1994)}]{Madden-Julian-1994:observations}
Madden, R.~A., and P.~R. Julian (1994), Observations of the 40-50-day tropical
  oscillation- a review, \textit{Mon. Weath. Rev.}, \textit{122}, 814--837.

\bibitem[{\textit{Maloney and Sobel}(2004)}]{Maloney-Sobel-2004:surface}
Maloney, E.~D., and A.~H. Sobel (2004), Surface fluxes and ocean coupling in
  the tropical intraseasonal oscillation, \textit{J. Climate}, \textit{17},
  4368Ð4386.

\bibitem[{\textit{Marshall et~al.}(1997{\natexlab{a}})\textit{Marshall,
  Adcroft, Hill, Perelman, and Heisey}}]{Marshall-et-al-1997:finite}
Marshall, J., A.~Adcroft, C.~Hill, L.~Perelman, and C.~Heisey
  (1997{\natexlab{a}}), A finite-volume, incompressible navier stokes model for
  studies of the ocean on parallel computers, \textit{J. Geophys. Res.},
  \textit{102, C3}, 5,753--5,766, mars-eta:97b.

\bibitem[{\textit{Marshall et~al.}(1997{\natexlab{b}})\textit{Marshall, Hill,
  Perelman, and Adcroft}}]{Marshall-et-al:1997:hydrostatic}
Marshall, J., C.~Hill, L.~Perelman, and A.~Adcroft (1997{\natexlab{b}}),
  Hydrostatic, quasi-hydrostatic and nonhydrostatic ocean modeling, \textit{J.
  Geophys. Res.}, \textit{102, C3}, 5,733--5,752, mars-eta:97a.

\bibitem[{\textit{Neelin et~al.}(1987)\textit{Neelin, Held, and
  Cook}}]{Neelin-Held-Cook-1987:evaporation}
Neelin, J.~D., I.~M. Held, and K.~H. Cook (1987), Evaporation-wind feedback and
  low-frequency variability in the tropical atmosphere, \textit{J. Atmos.
  Sci.}, \textit{44}, 2341--2348.

\bibitem[{\textit{Raymond}(2001)}]{Raymond-2001:new}
Raymond, D.~J. (2001), A new model of the {Madden-Julian} oscillation,
  \textit{J. Atmos. Sci.}, \textit{58}(18), 2807--2819.

\bibitem[{\textit{Sobel et~al.}(2008)\textit{Sobel, Maloney, Bellon, and
  Frierson}}]{Sobel-Maloney-Bellon-Frierson-2008:role}
Sobel, A.~H., E.~D. Maloney, G.~Bellon, and D.~Frierson (2008), The role of
  surface heat fluxes in tropical intraseasonal oscillations, \textit{Nature
  Geosicence}, \textit{1}, 653--657.

\bibitem[{\textit{Solodoch}(2010)}]{Solodoch-2010:excitation}
Solodoch, A. (2010), Excitation of slow kelvin waves in the equatorial
  atmosphere by yanai wave-group-induced convection, Master's thesis, Weizmann
  Institute of Science.

\bibitem[{\textit{Solodoch et~al.}(2010)\textit{Solodoch, Boos, Kuang, and
  Tziperman}}]{Solodoch-Boos-Kuang-et-al-2010:mjo}
Solodoch, A., W.~Boos, Z.~Kuang, and E.~Tziperman (2010), Mjo-like signal in
  due to yanai-wave forcing in intermediate-complexity atmospheric models,
  \textit{in prep}.

\bibitem[{\textit{Wheeler and
  Kiladis}(1999)}]{Wheeler-Kiladis-1999:convectively}
Wheeler, M., and G.~N. Kiladis (1999), Convectively coupled equatorial waves:
  Analysis of clouds and temperature in the wavenumber-frequency domain,
  \textit{J. Atmos. Sci.}, \textit{56}(3), 374--399.

\bibitem[{\textit{Yang and Ingersoll}(2009)}]{Yang-Ingersoll-2009:testing}
Yang, D., and A.~P. Ingersoll (2009), Testing a rossby-wave theory of the mjo,
  in \textit{AGU fall meeting abstract}.

\bibitem[{\textit{Zhang}(2005)}]{Zhang-2005:MJOrev}
Zhang, C.~D. (2005), Madden-julian oscillation, \textit{Rev. Geophys.},
  \textit{43}.

\end{thebibliography}
\end{document}